\documentclass[aps,prl,showpacs,twocolumn,groupedaddress]{revtex4}

\usepackage{graphics}

\begin{document}

\title{Landau levels and the Thomas-Fermi structure of rapidly
rotating Bose-Einstein condensates}

\author{Gentaro Watanabe,$^{a,b,c}$ Gordon Baym,$^{a,b,d}$ and C. J.
Pethick$^{b}$}
\affiliation{
$^{a}$Department of Physics, University of Tokyo,
Tokyo 113-0033, Japan
\\
$^{b}$NORDITA, Blegdamsvej 17, DK-2100 Copenhagen \O, Denmark
\\
$^{c}$Computational Astrophysics Laboratory, RIKEN,
Saitama 351-0198, Japan
\\
$^{d}$Department of Physics, University of Illinois, 1110 W. Green St.,
Urbana, IL 61801}

\date{\today}

\begin{abstract}

We show that, within mean-field theory, the density profile of a
rapidly rotating harmonically trapped Bose-Einstein condensate is of
the Thomas-Fermi form as long as the number of vortices is much larger
than unity. Two forms of the condensate wave function are explored:  i)
the lowest Landau level (LLL) wave function with a regular lattice of
vortices multiplied by a slowly varying envelope function, which gives
rise to components in higher Landau levels; ii) the LLL wave function with
a nonuniform vortex lattice. From variational calculations we find it most
favorable energetically to retain the LLL form of the wave function but to
allow the vortices to deviate slightly from a regular lattice. The
predicted distortions of the lattice are small, but in accord with recent
measurements at lower rates of rotation.

\end{abstract}

\pacs{03.75.Hh, 05.30.Jp, 67.40.Vs, 67.40.Db}

\maketitle

    How very rapidly rotating Bose-Einstein condensates carry angular momentum
remains a fundamental question of many-particle physics.  Ho \cite{ho}, noting
that the Hamiltonian for a rotating gas in a harmonic trap is similar to that
for charged particles in a magnetic field, argued that for rotational angular
velocities just below the transverse trap frequency, $\omega_{\perp}$, all
particles would condense into the lowest Landau level (LLL) of the Coriolis
force.  Motivated by this insight, Schweikhard {\it et al.}  \cite{profile}
(following earlier work in \cite{lattice2}) have recently achieved rotational
angular velocities $\Omega$ in excess of $0.99\,\omega_{\perp}$, at which the
cloud contains several hundred vortices.  The experiments have reached the
``mean-field quantum Hall'' frontier at which $\hbar \Omega$ becomes
comparable to the interaction energy, $gn$, where $n$ is the particle density,
$g=4\pi \hbar^2a_{\rm s}/m$ is the two-body interaction strength, $m$ is the
particle mass, and $a_{\rm s}$ the s-wave scattering length.

    Employing a quantum-Hall like condensate wave function with only LLL
components,
\begin{equation}
  \phi_{\rm LLL}({\bf r})=A_{\phi}
 \prod_{i=1}^{N}(\zeta-\zeta_i)\ e^{-r^2/2d_{\perp}^2},
 \label{LLL}
\end{equation}
-- where the rotation axis is along $\hat z$, $\zeta=x+iy$, the $\zeta_i$
are the vortex positions, ${\bf r} =(x,y)$, $d_{\perp}\equiv \sqrt{\hbar/m\omega_{\perp}}$ is the transverse 
oscillator length, and the
constant $A_{\phi}$ normalizes $\int d^2r |\phi_{\rm LLL}|^2$ to unity -- Ho
predicted that for a uniform lattice (at each height), the
smoothed density profile of a trapped cloud would be Gaussian:
\begin{equation}
  \langle |\phi_{\rm LLL}^{\rm u}|^2 \rangle
  = \frac{1}{\pi \sigma^2} e^{-r^2/\sigma^2}.
\end{equation}
Here the superscript ``u" denotes that the vortex density is uniform,
$1/\sigma^2=1/d_\perp^2-\pi n_{\rm v}$, $n_{\rm v}$ is the vortex density in
the plane perpendicular to the rotation axis, and $\langle \ldots \rangle$
denotes an average over an area of linear size large compared with the vortex
separation but small compared with $\sigma$ \cite{ho}.  On the other hand,
Ref.~\cite{bp} -- adopting a more general wave function, as in \cite{fb}, that
is a product of a slowly varying envelope describing the global structure of
the cloud and a rapidly varying function describing the local properties of
individual vortices -- found that for a uniform array of vortices and for
$Na_{\rm s}/Z \gg 1$, where $Z$ is the height in the $z$ direction, the
density profile is a Thomas-Fermi (TF) inverted parabola.  In current
experiments \cite{profile}, $Na_{\rm s}/Z$ is of order 10-100, and the
transverse density distributions are in fact better fit with a TF profile than
a Gaussian.  Theoretical evidence for the Thomas-Fermi profile has previously
been found in numerical studies \cite{sinova2}.

    As first pointed out by A.H.  MacDonald (see Ref.\ 37 of \cite{coddington})
and stressed to us by N. Read \cite{read}, 
one can achieve significant changes in the density while
maintaining a condensate wave function made only of LLL components, if one
allows the vortex lattice to relax from exactly triangular, an effect not considered in Refs.\ \cite{ho,bp}.  
We show here that
when the lattice is permitted to relax only slightly from uniformity, the TF
form is generally obtained for $Na_{\rm s}/Z \gg 1-\Omega/\omega_\perp$.  Thus
TF behavior extends to much lower densities than previously realized.  This
effect comes about through a modification, with increasing $gn$, of the
condensate wave function, $\Psi$, from $N^{1/2}\phi^{\rm u}_{\rm LLL}$.  A
modification of $\Psi$ that allows the density distribution to spread out
tends to reduce the interaction energy:  for infinitesimal modification, $\Psi
\sim \phi_{\rm LLL}^{\rm u} + \delta\Psi$, the change in the interaction
energy is linear in $\delta\Psi$, and, with the proper choice of the phase of
$\delta\Psi$, is negative.  Such modification costs kinetic and trap energy,
which however is quadratic in $\delta\Psi$ for infinitesimal modification.
Thus the state (\ref{LLL}) for a uniform array of vortices is always unstable
towards smoothing out of the density distribution.  The resulting change in
the density profile can be large even though the distortions of the lattice
are small.

    We consider the properties of a rotating cloud in a harmonic transverse
potential $V(r)=m\omega_\perp^2 (x^2+y^2)/2$.  For simplicity, we treat mainly
the two-dimensional problem, and generally set $\hbar = 1$. The approach adopted in Ref.\ \cite{bp}
was to express quantities as sums over Wigner-Seitz cells for single vortices.  This is cumbersome, 
and in this Letter we adopt a different trial
condensate wave function which is better suited for rotational angular velocities close to $\omega_{\perp}$,
\begin{equation}
\Psi({\bf r})= N^{1/2}h(r) \phi_{\rm LLL} \equiv N^{1/2}\psi({\bf r})\ .
\label{ansatz}
\end{equation}
The function $h$, which we assume to be real and slowly varying on the
scale of the intervortex separation, modifies the LLL components, as well as
admixes higher Landau levels. The vortex lattice need not be uniform. The wave function (\ref{ansatz})
is much more convenient for calculating the kinetic energy than is the ansatz used in Ref.\ \cite{bp}.   

The admixture $b_{\mu,\nu}$ of the higher Landau level $(\mu,\nu)$ in the
wave function (\ref{ansatz}), where $\mu$ is the angular momentum index and
$\nu$ the level index, is of order $d_\perp^\nu \left(d^\nu
h/dr^\nu\right)_{r_\mu} \sim (d_\perp/R)^\nu \left(h(0)-1\right)C_{\mu}^0$,
where $C_\mu^0$ is the amplitude of the level $\mu$ in $\phi_{\rm LLL}$, and
$r_\mu = d_\perp\sqrt{\mu}$ is the peak of the LLL wave function $\mu$
\cite{footnote1}.  The amplitude for admixture of higher Landau levels is thus
of order $d_{\perp}/R \sim 1/N_{\rm v}^{1/2}$ relative to the modification of
the LLL contribution, where $N_{\rm v}$ is the total number of vortices in the
cloud.

    The energy per particle of the condensate in the rotating frame is
$E'=E-\Omega L_z$, where $E$ is the energy in the non-rotating frame and
$L_z$ is the expectation value of the angular momentum per particle about the
rotation axis.  Following Ho \cite{ho}, we write
\begin{eqnarray}
   &E'& = (\omega_{\perp}- \Omega) L_z \nonumber\\
 &&\hspace{-1cm} + \int d^2r\ \psi^*
 \left[\frac{m}{2}\left( \frac{\nabla_{\perp}}{im}
 -\mbox{\boldmath $\omega$}_{\perp}\!\! \times \mbox{\boldmath $r$}\right)^2
 + \frac{g_{\rm 2D}}{2} |\psi|^2\right] \psi\ ,
 \label{k original}
\end{eqnarray}
where $\mbox{\boldmath $\omega$}_{\perp}= \omega_{\perp} \hat{\bf z}$, and
$g_{\rm 2D}$ is the effective coupling parameter in two dimensions.  If the
system is uniform in the $z$ direction, $g_{\rm 2D}=Ng/Z$, where $Z$ is the
axial extent of the cloud, while if the system in the $z$ direction is in the
ground state of a particle in a harmonic potential of frequency $\omega_{z}$,
then $g_{\rm 2D}=Ng/d_z\sqrt{2\pi}$, where $d_z\equiv(\hbar/m\omega_z)^{1/2}$
\cite{fetter2}.

    Because the higher Landau levels in the variational wave function
(\ref{ansatz}) have probability $\sim 1/N_{\rm v}$, the angular momentum per
particle is
\begin{equation}
  L_z = \int d^2r\, \frac{r^2}{d_{\perp}^2}h^2 |\phi_{\rm LLL}|^2 \, -1,
\end{equation}
plus terms of order $1/N_{\rm v}$, which we neglect.  Furthermore, since $h$
varies slowly in space, we shall use the averaged vortex approximation as in
\cite{ho} to write \cite{dndr}
\begin{eqnarray}
   E'\simeq \Omega +\int d^2r\ \langle |\phi_{\rm LLL}|^2 \rangle
  \left\{
  \frac{1}{2m}  \left(\frac{dh}{dr}\right)^2\right.\nonumber \\
  \left.\vphantom{\left(\frac{dh}{dr}\right)^2}
  +(\omega_{\perp}-\Omega) \frac{r^2}{d_{\perp}^2}  h^2
  +\frac{bg_{\rm 2D}}{2}\langle |\phi_{\rm LLL}|^2 \rangle h^4
    \right\}\ .
\label{k}
\end{eqnarray}
Here $b \equiv \langle |\phi_{\rm LLL}|^4 \rangle / \langle |\phi_{\rm
LLL}|^2 \rangle^2 \simeq 1.158$ describes the renormalization of the effective
interaction due to the rapid density variations on the scale of the vortex
separation \cite{fb,bp,sinova}.

{\it Uniform lattice.}    We begin with the ansatz (\ref{ansatz}) for a uniform vortex density.  The
first term in braces in Eq.~(\ref{k}), the extra kinetic energy due to
admixture of excited Landau levels, scales as $1/R^2$, as does the interaction
energy.  If $g_{\rm 2D} \ll \hbar^2/m$, the first term suppresses spatial
variations of $h$, and $h\approx 1$, i.e., the density profile is Gaussian.
This condition is simply that $Na_{\rm s}/Z \ll 1$, where $Z$ is the effective
size in the $z$ direction, bounded below by $d_z$.  In the opposite limit,
$g_{\rm 2D} \gg \hbar^2/m$, the extra kinetic energy is unimportant, and the
optimal density profile is obtained by minimizing the second and third terms
in the integrand, which results in a TF density profile, in agreement with the
considerations of Ref.\ \cite{bp}.

    To obtain illustrative results we employ a variational trial function that
interpolates between the Gaussian and TF (G-TF) forms; exploiting the fact that
$\lim_{\alpha\rightarrow\infty} (1-t/\alpha)^\alpha =e^{-t}$, as in
\cite{fetter}, we take
\begin{equation}
 h(r)= A_h \left(1 - \frac{r^2}{\alpha L^2}\right)^{\alpha/2}
 e^{r^2/2\sigma^2}\ ,\label{h gaussian-tf}
\end{equation}
for $0 \le r < \sqrt{\alpha}L$, and $h=0$ otherwise.  Expression (\ref{h
gaussian-tf}) describes both the Gaussian ($\alpha \to \infty$) and
TF ($\alpha =1$) regimes.  The number of particles per unit area is
$Nh^2 \langle |\phi^{\rm u}_{\rm LLL}|^2 \rangle$; thus
$A_h^2=(\sigma^2/L^2)(1+1/\alpha)$.  We refer to Eq.~(\ref{h gaussian-tf}) for
$h$ as the G-TF form.

    Substituting Eq.\ (\ref{h gaussian-tf}) into (\ref{k}), we obtain
\begin{eqnarray}
 E'_{\rm G{\mbox -}TF} &=&
 \Omega + \frac{\omega_{\perp}}{2} d_{\perp}^2
 \left( \frac{1}{L^2}\frac{\alpha+1}{\alpha-1} - \frac{2}{\sigma^2}
 + \frac{\alpha}{\alpha+2}\frac{L^2}{\sigma^4} \right) \nonumber\\
 && + (\omega_{\perp}-\Omega)
 \frac{\alpha}{\alpha+2}\frac{L^2}{d_{\perp}^2}
 +\frac{b g_{\rm 2D}}{2\pi}
 \frac{1}{L^2}\frac{(\alpha+1)^2}{\alpha(2\alpha+1)}\ .
\nonumber\label{energy g-tf}\\
\end{eqnarray}
Minimization of $E'_{\rm G{\mbox -}TF}$ with respect to $\sigma$ and $L$
yields
\begin{eqnarray}
 L^2 &=& d_{\perp}^2
 \left(1-\frac{\Omega}{\omega_{\perp}}\right)^{-\frac{1}{2}} \nonumber\\
 && \times
 \frac{1}{\alpha}\left[
 (\alpha+2) \left\{ \frac{1}{\alpha-1}
 + \kappa \frac{(\alpha+1)^2}{(2\alpha+1)}
 \right\} \right]^{\frac{1}{2}} ,\label{lsq}
\end{eqnarray}
and
\begin{eqnarray}
  \sigma^2 &=& L^2/(1+2/\alpha)
 \ ,\label{sigmasq}
\end{eqnarray}
where the dimensionless parameter
%\begin{equation}
$
\kappa \equiv m b g_{\rm 2D}/(2\pi\hbar^2)\
%\kappa \equiv \frac{m b g_{\rm 2D}}{2\pi\hbar^2}\
$
%\label{kappa}
%\end{equation}
determines the strength of interparticle interactions.

    The optimal value of $\alpha$, determined by minimizing $E'_{\rm G{\mbox
-}TF}$ with respect to $\alpha$, obeys the quartic equation, $\alpha^4
-2(1+4\lambda) \alpha^3 -12 \lambda\alpha^2 +2(1-3\lambda)\alpha
-1-\lambda=0$, where $\lambda\equiv\kappa^{-1}$.  The only real and positive
solution is
\begin{eqnarray}
  2\alpha &=& \left(2+32\eta + 3\cdot 2^{\frac23} \eta^{\frac13}
 +\frac{2(1+2\lambda)(32\eta -1)}
  {\sqrt{1+16\eta -3\cdot 2^{\frac23}\eta^{\frac{1}{3}}}} \right)^{\frac12}
  \nonumber\\
  &&+
  (1 + 4\lambda) +\sqrt{1+16\eta-3\cdot 2^{\frac{2}{3}}
  \eta^{\frac{1}{3}}} ,
  \label{alpha}
\end{eqnarray}
where $\eta = \lambda(1+\lambda)$.  We note that $\alpha$ is independent
of $\Omega$.  In the weak $(\kappa\to 0)$ and strong interaction
$(\kappa\to \infty)$ limits,
\begin{eqnarray}
\alpha &\simeq & 8/\kappa \qquad {\mbox (\kappa\ll 1)}\ ,
 \label{alpha weak int}\\
 \alpha &\simeq& 1+(3/2^\frac13) \kappa^{-\frac{1}{3}}
 +{\cal O}(\kappa^{-\frac{2}{3}})
 \qquad {\mbox (\kappa^\frac{1}{3}\gg 1)}\label{alpha strong int}\ .
\end{eqnarray}

    The shape index $\alpha$, Eq.\ (\ref{alpha})
%, plotted in Fig.\ \ref{figalpha}, 
decreases from infinity in the absence of interaction to unity as
$\kappa$ increases from zero to infinity, and the density profile of the cloud
changes from a Gaussian to an inverted parabola.  Without the minimization
with respect to $\alpha$, we can describe the cloud by assuming a shape of the
density profile corresponding to a given value of the shape index, in terms of
which $L$ and $\sigma$ are given by Eqs.\ (\ref{lsq}) and (\ref{sigmasq}).  If
we let $\alpha\to\infty$, we reproduce a Gaussian density profile as in the
case without the modulating function, but with optimized width.

%\begin{figure}[htbp]
%\begin{center}\vspace{0.0cm}
%\rotatebox{0}{\hspace{-0.cm}
%\resizebox{7.5cm}{!}
%{\includegraphics{alpha_0to0.1_new1.eps}}}
%\caption{\label{fig alpha}
%    Shape index $\alpha$ as a function of $\kappa^{-1}$ over the range of
%interaction strengths typical of current experiments ($\kappa\simeq 10^2$).
%In the inset, $\alpha$ is plotted over a wider range of $\kappa^{-1}$ to show
%its asymptotic behavior (\ref{alpha weak int}).
%  }
%\end{center}
%\end{figure}

\begin{figure}[htbp]
\begin{center}\vspace{0.0cm}
\rotatebox{0}{\hspace{-0.cm}
\resizebox{8.0cm}{!}
{\includegraphics{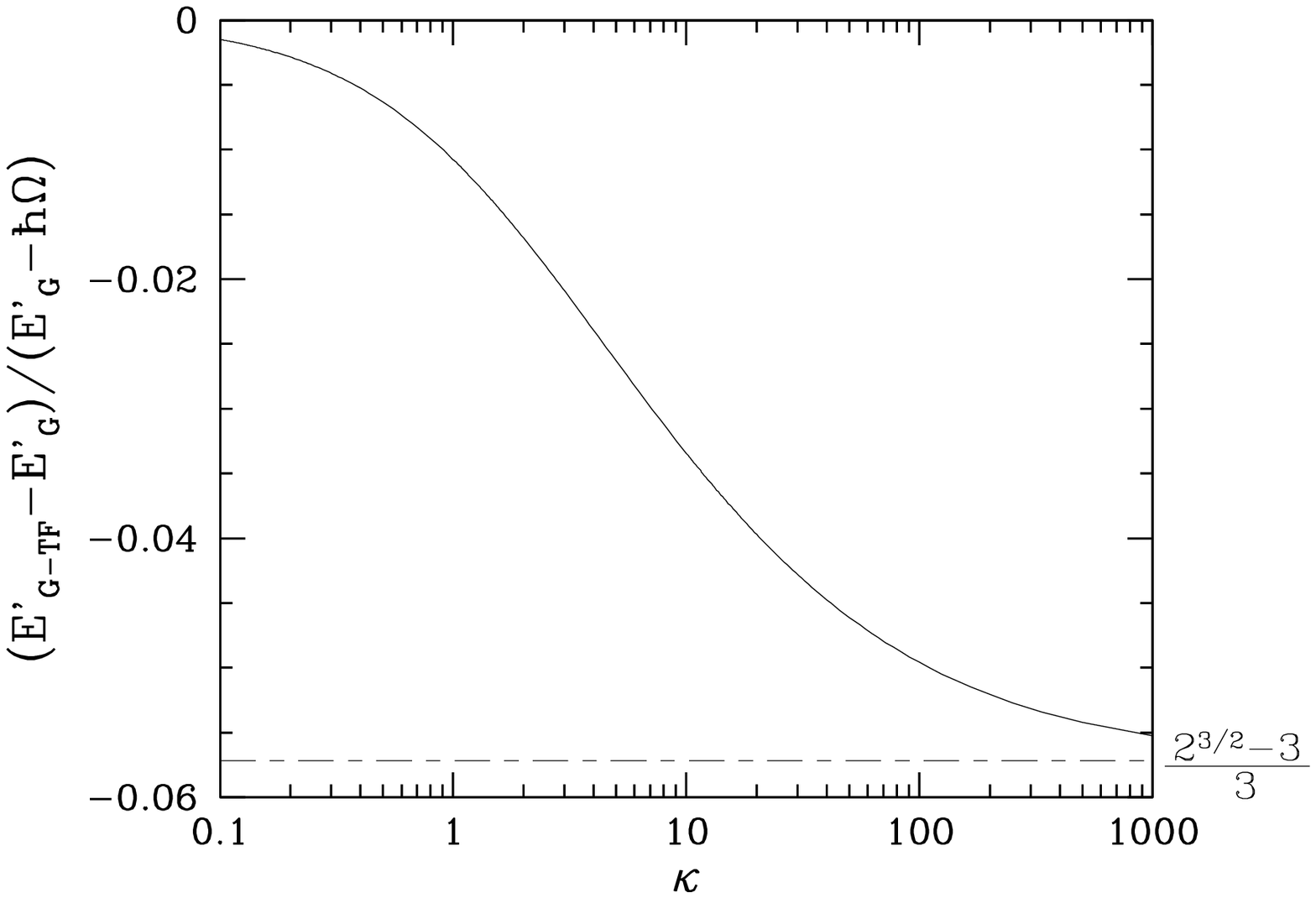}}}
\caption{\label{fig energy}
    Difference between the energies $E'_{\rm G{\mbox -}TF}$ calculated in G-TF
with optimized $\alpha$, Eq.\ (\ref{alpha}), and that for the Gaussian case
with $\alpha\to \infty$, measured relative to $E'_{\rm G}-\Omega$.  The
curve converges to $(2^{3/2}-3)/3 \simeq -0.0572$ as $\kappa\to \infty$.
  }
\end{center}
%\end{figure}
%
%
%\begin{figure}[htbp]
%\begin{center}
\vspace{0.2cm}
\rotatebox{0}{\hspace{-0.cm}
\resizebox{8.0cm}{!}
{\includegraphics{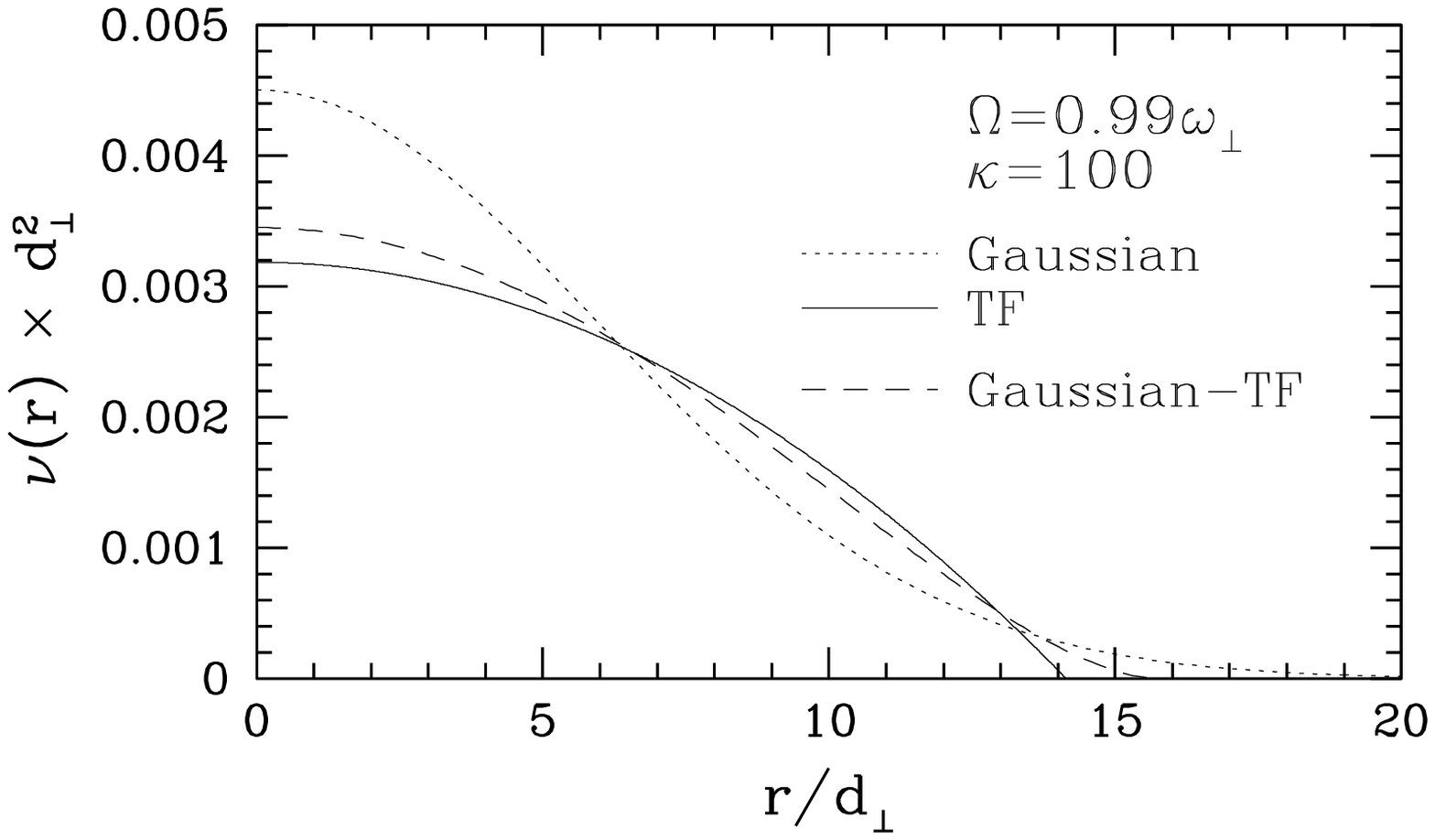}}}
%{\includegraphics{profile_compare_omg0.99_k0.01_negl2nd_rev1_new.eps}}}
\caption{\label{fig profile compare} Global density profile $\nu(r)=h^2
\langle |\phi_{\rm LLL}|^2 \rangle$ of the cloud for
$\Omega=0.99\,\omega_{\perp}$ and $\kappa=100$.  The dashed line is for the
optimized $\alpha$ given by Eq.\ (\ref{alpha}), and the dotted line is the
Gaussian case.  The solid line is the TF profile, Eq.\ (\ref{tfprofile}).
  }
%\end{center}
\end{figure}

    We now compare the results for the G-TF profile with $\alpha$ optimized
and the profile with $\alpha\rightarrow\infty$.  We write the energy for the
Gaussian profile as
\begin{eqnarray}
  E'_{\rm G} &\simeq& \Omega + \sqrt{2}\ \omega_{\perp}
  \left(1-\frac{\Omega}{\omega_{\perp}}\right)^{\frac{1}{2}}
  \kappa^{\frac{1}{2}}\ ,
\end{eqnarray}
obtained by taking $\alpha\to\infty$ in Eq.\ (\ref{energy g-tf}).
The relative energy difference
$(E'_{\rm G{\mbox -}TF}-E'_{\rm G})/(E'_{\rm G}- \Omega)$
between the G-TF and Gaussian cases is shown
in Fig.\ \ref{fig energy}.
In the limit $\kappa \rightarrow \infty$, the energy is the TF result,
\begin{eqnarray}
 E'_{\rm TF} \simeq \Omega
 + \frac{4}{3}\omega_{\perp}
 \left(1-\frac{\Omega}{\omega_{\perp}}\right)^{\frac{1}{2}}
 \kappa^{\frac{1}{2}},
\label{TF}
\end{eqnarray}
and $(E'_{\rm G{\mbox -}TF}-E'_{\rm G})/(E'_{\rm G} -\Omega)$
converges to $(2^{3/2}/3)-1\simeq -0.0572$, independent of $\Omega$ (Fig.\
\ref{fig energy}).

    In the experiments in Ref.~\cite{profile}, where $\omega_z=2\pi\times5.3$
Hz, $N=1.5\times10^5$, and $a_{\rm s}=5.6$ nm for the triplet state of
$^{87}$Rb, one has $Na_{\rm s}/Z \agt 26$ at $\Omega=0.989\, \omega_{\perp}$,
typical of the highest angular velocities achieved so far.  For a uniform
density in the axial direction, with $g_{\rm 2D}=g N/Z$, we find
$\kappa=2Na_{\rm s}b/Z \agt 52$, while for a Gaussian density profile in $z$,
for which $g_{\rm 2D}=g N/d_z\sqrt{2\pi}$, we obtain $\kappa=2Na_{\rm s}b/
d_z\sqrt{2\pi} \simeq 1.4\times 10^2$.

    In Fig.\ \ref{fig profile compare} we plot the normalized density profile
$\nu(r)=h^2 \langle |\phi_{\rm LLL}|^2 \rangle$ at
$\Omega=0.99\,\omega_{\perp}$ and $\kappa=100$ for the G-TF case in addition
to $\nu(r)$ for the two extreme cases of the Gaussian form with
$\alpha\rightarrow\infty$ and the inverted TF parabola; see below.  The
density profile for the G-TF modulation (dashed line) is closer to the TF form
than to a Gaussian.  Figure~\ref{fig energy} indicates that the relative
energy reduction compared with $E-\Omega$ for the Gaussian approximation is
only $\simeq 4.95\%$.

{\it Distorted lattice.}    Next we consider an LLL wave function with a non-uniform vortex density,
and take $h(r)=1$.  Equation (\ref{k}) is still valid, and the energy, given
by
\begin{equation}
   E'= \Omega
   +\int d^2r\{(\omega_{\perp}-\Omega) \frac{r^2}{d_{\perp}^2}
   \langle |\phi_{\rm LLL}|^2\rangle
  +\frac{bg_{\rm 2D}}{2}\langle |\phi_{\rm LLL}|^2 \rangle^2\},
\label{k2}
\end{equation}
is minimized by the TF profile
\begin{equation}
  \langle |\phi_{\rm LLL}|^2 \rangle =
  \frac{1}{bg_{\rm 2D}}
  \left(\mu-\Omega -(\omega_{\perp}-\Omega) \frac{r^2}{d_{\perp}^2}\right),\label{tfprofile}
\end{equation}
where $\mu$ is the chemical potential.  Such a solution is possible only
if one relaxes the constraint of a regular lattice; otherwise the solution is
a Gaussian, as in \cite{ho}.  The energy is given by the TF result (\ref{TF}).  Since the term in Eq.\ (\ref{k}) 
involving $dh/dr$ is positive definite, it is clear that it is energetically favorable to create deviations of the density profile from $|\phi^{\rm u}_{\rm LLL}|^2$ by deforming the lattice rather than by exciting higher Landau levels.  
These conclusions have recently been confirmed by numerical calculations by Cooper {\it et al.} \cite{cooper}. 
Excitation of higher Landau levels will reduce the energy of a distorted lattice still further, but it may be shown that this is of order
 $(gn)^2/\hbar \Omega$ per particle, which is smaller than the terms we have retained. 

Remarkably, the energy and TF profile are valid for all interactions strong
enough that the averaged vortex approximation is valid.  This holds provided
the size of the cloud is large compared with $d_{\perp}$, or alternatively
that the number of vortices in the cloud $N_{\rm v}\approx R^2/d_{\perp}^2$ is
large compared with unity.  This condition is equivalent to $Na_{\rm s}/Z \gg
1-\Omega/\omega_\perp$, i.e., that the energy, $\hbar(\omega_\perp-\Omega)$,
to excite higher angular momenta $\mu$ in the LLL, be small compared with the
mean field energy in the situation that all particles are in the transverse
ground state of the trap.  For $\Omega$ close to $\omega_\perp$, this
condition for a TF profile is much weaker than that for a non-rotating cloud.
By contrast, a TF profile with a uniform lattice requires $Na_{\rm s}/Z \gg
1$.

    Modifying the amplitudes of lowest Landau level components in the
condensate wave function, without introducing excited Landau level components,
is equivalent to changing the positions of the vortices.  In the LLL wave
function (\ref{LLL}), the positions of vortices determine the smoothed density
distribution generally through
\cite{ho},
\begin{equation}
 \frac{1}{4}\nabla^2 \ln n(r)  = -\frac{1}{ d_\perp^2} +\pi n_{\rm v}(r).
  \label{nv}
\end{equation}
This equation allows us to estimate the displacement of the vortices from
a triangular lattice for a general density distribution.  In particular, if we
assume a TF distribution, $n(r)\sim 1-r^2/R^2$, where $R$ is the radial extent
of the cloud, then Eq.~(\ref{nv}) implies \cite{sheehy0}
\begin{equation}
   n_{\rm v}(r) = \frac{1}{\pi d_\perp^2} -
  \frac{1}{\pi R^2}\frac{1}{\left(1-r^2/R^2\right)^2}.
  \label{nnv}
\end{equation}
The second term is of order $1/N_{\rm v}$ compared with the first, since
$N_{\rm v}\simeq R^2/d_\perp^2$.  The corresponding mean displacement $\delta
r$ in the radial direction of the vortices at radius $r$ is then $\delta r/r =
(d_\perp^2/2R^2)/(1-r^2/R^2) \sim 1/N_{\rm v}$.

    Thus in the LLL limit very small distortions of the lattice can result in
large changes in the density distribution.  For lower rotation rates, Sheehy
and Radzihovsky \cite{sheehy} have recently demonstrated that the vortex
density obeys an equation of the form (\ref{nnv}) but with a coefficient of
the second term which depends on the ratio of the vortex separation to the
vortex core size.  Such distortions of the vortex lattice have been
measured experimentally at relatively low rotation rates \cite{coddington},
and are in good agreement with theory \cite{sheehy}.  It would be valuable to extend the
measurements to rapidly rotating condensates.

    We thank Nick Read for stimulating correspondence, and V. Cheianov, Nigel Cooper and
James Anglin for helpful comments.  Author GW thanks K. Sato, K. Yasuoka, and
T. Ebisuzaki for encouragement and L. M. Jensen, and P. Urkedal for help with
computer facilities.  Author GB thanks the Japan Society for the Promotion of
Science and T. Hatsuda for enabling him to spend a fruitful research period at
the University of Tokyo.  This work was supported in part by Grants-in-Aid for
Scientific Research provided by the Ministry of Education, Culture, Sports,
Science and Technology through Research Grant No. 14-7939, by the Nishina
Memorial Foundation, and by NSF Grant PHY00-98353.


\begin{thebibliography}{99}
\bibitem{footnote1} The amplitude of a higher Landau level $(\mu,\nu)$ in
the wave function (\ref{ansatz}) is $b_{\mu,\nu}=\int d^2r
\chi_{\mu,\nu}^*h(r)\chi_{\mu,0}C_{\mu}^0$, where $\chi_{\mu,\nu}$ is the wave
function of the level $(\mu,\nu)$. $C_\mu^0$ is the amplitude of the level
$\mu$ in $\phi_{\rm LLL}$.  Since $\chi_{\mu,0}$ is strongly peaked about
$r_\mu =d_\perp\sqrt{\mu}$, $b_{\mu,\nu} \sim d_\perp^\nu \left(d^\nu
h/dr^\nu \right)_{r_\mu} \sim (d_\perp/R)^\nu \left(h(0)-1\right)C_{\mu}^0$.
Further details will be given in a future publication.
\bibitem{dndr} Since the smoothed density profile $n(r)$ equals
$Nh^2|\langle\phi_{\rm LLL}^{\rm u}\rangle|^2$, one readily derives the
equivalence, to order $1/N_{\rm v}^2$, of Eq.~(\ref{k}) and the transverse
terms of Eq.~(36) of \cite{bp}.
\end{thebibliography}
\end{document}